\documentclass[10pt]{article}

\usepackage{amsmath}
\usepackage{amssymb}
\usepackage{amsmath,amsthm,amssymb,amscd}
\usepackage{latexsym}
\usepackage{indentfirst}
\usepackage{subfigure}
\usepackage{graphicx}
\usepackage{extarrows}
\usepackage{bm}
\usepackage{multirow}
\usepackage[colorlinks,linkcolor=blue,hyperindex,dvipdfm]{hyperref}

\usepackage{cite}
\usepackage{color}

\usepackage[labelfont=bf,labelsep=period,justification=raggedright]{caption}

\makeatletter
\renewcommand{\@biblabel}[1]{\quad#1.}

\renewcommand\thetable{\arabic{table}}

\makeatother

\date{}

\begin{document}

% Title must be 150 characters or less
\begin{flushleft}
{\Large
\textbf{Cargo transportation by two species of motor protein}
}
% Insert Author names, affiliations and corresponding author email.
\\
Yunxin Zhang$^{1,\ast}$,
\\
{1} Shanghai Key Laboratory for Contemporary Applied Mathematics, Laboratory of Mathematics for Nonlinear Science, Centre for Computational Systems Biology, School of Mathematical Sciences, Fudan University, Shanghai 200433, China.
\\
$\ast$ E-mail: xyz@fudan.edu.cn
\end{flushleft}

% Please keep the abstract between 250 and 300 words

\section*{Abstract}
The cargo motion in living cells transported by two species of motor protein with different intrinsic directionality is discussed in this study. Similar to single motor movement, cargo steps forward and backward along microtubule stochastically. Recent experiments found that, cargo transportation by two motor species has a memory, it does not change its direction as frequently as expected, which means that its forward and backward step rates depends on its previous motion trajectory. By assuming cargo has only the least memory, i.e. its step direction depends only on the direction of its last step, two cases of cargo motion are detailed analyzed in this study: {\bf (I)} cargo motion under constant external load; and {\bf (II)} cargo motion in one fixed optical trap. Due to the existence of memory, for the first case, cargo can keep moving in the same direction for a long distance. For the second case, the cargo will oscillate in the trap. The oscillation period decreases and the oscillation amplitude increases with the motor forward step rates, but both of them decrease with the trap stiffness.
The most likely location of cargo, where the probability of finding the oscillated cargo is maximum, may be the same as or may be different with the trap center, which depends on the step rates of the two motor species. Meanwhile, if motors are robust, i.e. their forward to backward step rate ratios are high, there may be two such most likely locations, located on the two sides of the trap center respectively. The probability of finding cargo in given location, the probability of cargo in forward/backward motion state, and various mean first passage times of cargo to give location or given state are also analyzed.

%\section*{Author Summary}
%One biophysical model is presented to study cargo motion in living cells by two species of motor protein. In this model, the cargo is assumed to have one-step memory, it is more likely to move forward/backward if it got to its present location by one forward/backward step. Two special cases are mainly discussed, {\bf (I)} cargo motion under constant external load; and {\bf (II)} cargo motion in one fixed optical trap.

%\keywords{cargo memory; tug-of-war; motor protein}
%
%\maketitle

\section*{Introduction}\label{introduction}
Motility is one of the basic properties of living cells, in which cargos, including organelles and vesicles, are usually transported by cooperation of various motor proteins \cite{Bray2001,Howard2001}, such as the plus-end directed kinesin and minus-directed dynein \cite{Block1990,Vale2003,Mallik2004}. Experiments found that, using the energy released in ATP hydrolysis \cite{Hua1997,Schnitzer1997,Coy1999,Gennerich2007}, these motors can move processively along microtubule with step size 8 nm and in hand-over-hand manner \cite{Asbury2003,Toba2006,Guydosh2009}.

Although numerous experimental and theoretical studies have been done to understand this cargo transportation process, so far the mechanism of which is not fully clear. In \cite{Lipowsky2005}, one basic model is presented by assuming cargo is transported by only one motor species and all the motors share the external load equally. Then in \cite{Lipowsky2008}, one more realistic tug-of-war model is designed, in which the cargo is assumed to be transported by two motor species with opposite intrinsic directionality, and motors can reverse their motion direction under large external load. According to some experimental phenomena this tug-of-war model seems reasonable \cite{Gennerich2006,Soppina2009}. In either of the models given in \cite{Lipowsky2005,Lipowsky2008}, the only interaction among different motors is that, motors from the same species share load equally and motors from different species act as load to each other. In \cite{Kunwar2008,Kunwar2010,Zhang20114}, some complicated models are presented, in which interactions among motors are described by linear springs. Recent experiments found that the tug-of-war model might not be reasonable enough to explain some experimental phenomena, so several new models are designed to try to understand the mechanism of cargo motion by multiple motors \cite{Rogers2009,Driver2010,Driver2011,Jamison2011,Uppulury2012,Kunwar2011,Bouzat2012}. Finally, more discussion about cargo transportation in cells can be found in \cite{Frank1995,Badoual2002,Adachi2007,Bieling2008,Mallik2009,Brouhard2010,Welte2010,Hendricks2010,Schroeder2010}.

In recent experiment \cite{Leidel2012}, by measuring cargo dynamics in optical trap, Leidel {\it et al.} found cargo motion along microtubule has memory. Cargo is more likely to resume motion in the same direction rather than the opposite one. This finding implies that, cargo location in the next time depends not only on its present location but also on how it reaches the present location. The behavior of cargo depends on its motion trajectory, which is different from the assumptions in previous models.
% in which cargo behavior depends only on its present location and the number of motor bound to .
In this study, one model for cargo motion with memory will be presented. But for simplicity, we assume that the cargo has only a little memory, it can only remember the motion direction in its last step. %The cargo at location $n$ can only know it previous location is $n-1$ or $n+1$, but cannot remember how it get to location $n-1$ or $n+1$.
The description and theoretical analysis of the model with memory will be first given in the next section, and then corresponding results will be presented in the following section. Results will be summarized in the final section.

\section*{Model for cargo motion with memory}\label{model}
In this study, the cargo is assumed to be tightly bound by two motor species: {\it plus-end} (or {\it forward}) motors and {\it minus-end} (or {\it backward}) motors. The forward and backward step rates of each {\it plus-end} motor are $u$ and $w$, and the forward and backward step rates of each {\it minus-end} motor are $f$ and $b$. Obviously $u\gg w$ but $b\gg f$ when the external load is low, since the intrinsic directionalities of the two motor species are opposite to each other, and the intrinsic motion direction of {\it plus-end} motor is plus-end directed (i.e. to the plus-end of microtubule), but the intrinsic motion direction of {\it minus-end} motor is minus-end directed (i.e. to the minus-end of microtubule).  %forward/backward motion of {\it plus-end} motor means the motor moves to the plus/minus-end of microtubule, but the forward/backward motion of {\it minus-end} motor means the motor moves to the minus/plus-end of microtubule, since the intrinsic directionality of the two motor species are opposite to each other.
By assuming that all motors from the same motor species share the load equally, we only need to discuss the simplest cases in which the cargo is transported by only one {\it plus-end} motor and one {\it minus-end} motor. For example, if there are $k$ {\it plus-end} motors, the total external load is $F_c$, the forward and backward step rates of one single {\it plus-end} motor are $u_c$ and $w_c$, and the motor step size is $l_c$. Then these $k$ {\it plus-end} motors can be effectively replaced by one single {\it plus-end} motor with load $F=F_c/k$, step rates $u=ku_c$ and $w=kw_c$, and step size $l_0=l_c/k$. %Similar, we assume there is only one minus-end motor.
Since the experiments in \cite{Leidel2012} showed that, the number of motors moving the cargo is usually the same in both directions, this study also assumes the step sizes of the {\it plus-end} motor and {\it minus-end} motor are the same (note, the step size of single {\it plus-end} motor kinesin and step size of single {\it minus-end} motor dynein are the same $l_0\approx 8$ nm \cite{Howard2001,Gennerich2007,Guydosh2009}).

This study will mainly discuss two special cases: {\bf (I)} Cargo moves under constant external load. {\it In vitro}, this constant load may be applied by one feedback optical trap, or {\it In vivo}, this constant load may be from the viscous environment with invariable drag coefficient. {\bf (II)} Cargo moves in one fixed optical trap, this case is easy to be performed experimentally, and so the corresponding theoretical results are easy to be verified.

\subsection*{Cargo Motion under constant load}\label{IIA}
\begin{figure}
\begin{center}
  \includegraphics[width=450pt]{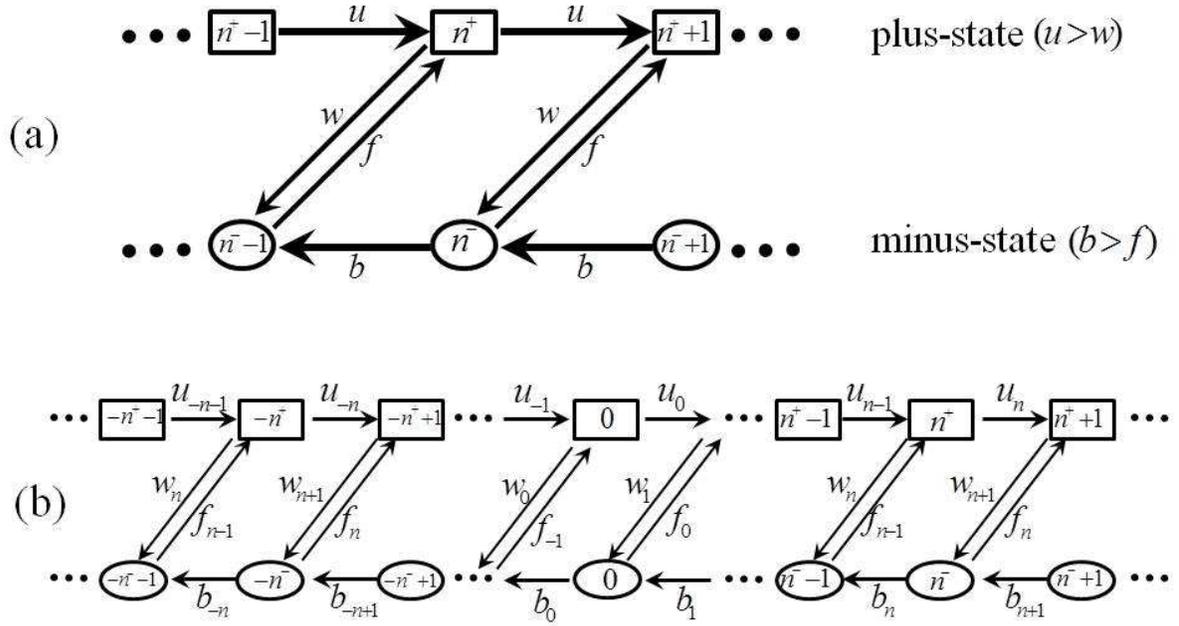}\\
\end{center}
  \caption{Schematic depiction of the model discussed in this study to explain the cargo motion with memory. (a) is for cargo motion under constant load, and (b) is for cargo motion in one fixed optical trap. At any location $n$, the cargo may be in two different states, {\it plus-state} $n^+$ and {\it minus-state} $n^-$. Cargo in {\it plus-state} $n^+$ means it reaches location $n$ from location $n-1$, while cargo in {\it minus-state} means it is from location $n+1$. For a cargo in {\it plus-state} $n^+$, its forward and backward step rates are $u$ and $w$ respectively. But for a cargo in {\it minus-state} $n^-$, it has different step rates $f$ and $b$. For the constant load cases (a), $u>w$ and $b>f$ mean that, if the cargo is in {\it plus-state} $n^+$ it will be more likely to move forward to location $n+1$. Otherwise, it will be more likely to move backward to location $n-1$. }\label{FigSchemtic}
\end{figure}

For the sake of convenience, the cargo is said to be in {\it plus-state} $n^+$ if it reached its present location $n$ by one forward step from location $n-1$. Similarly, the cargo is said to be in minus-state $n^-$ if its previous step is minus-end directed, see Fig. \ref{FigSchemtic}(a) for the schematic depiction. %This special cases can be schematically depicted by Fig. \ref{FigSchemtic}(a), Where $n^+, n^-$ denote that the cargo is at location $n$ but in plus-state and minus-state, respectively.
In {\it plus-state}, the forward step rate is higher than backward step rate $u>w$, but in {\it minus-state} the forward step rate is lower than backward step rate $f<b$. So in {\it plus-state}, the cargo is more likely to move forward, but in {\it minus-state}, the cargo will be more likely to move backward.
For example, for a cargo in location $n$, if its previous step is plus-end directed, from either {\it plus-state} $n^{+}-1$ or {\it minus-state} $n^--1$ to location $n$, then in the next step the cargo will be more likely to move to location $n+1$ ({\it plus-state} $n^++1$), since the cargo is now in plus-state $n^+$ and its forward step rate $u$ is higher than its backward step rate $w$. On the contrary, if it got to its present location $n$ from location $n+1$ (either from {\it plus-state} $n^++1$ or from {\it minus-state} $n_-+1$), then in the next step the cargo will be more likely to move to location $n-1$ ({\it minus-state} $n^--1$), since the cargo is now in {\it minus-state} $n^-$ and its backward step rate $b$ is higher than its forward step rate $f$. This behavior means that the cargo can remember its motion direction of its last step.

Let $p, \rho$ be probabilities of cargo in {\it plus-state} and {\it minus-state} respectively, then
\begin{equation}\label{eq1}
dp/dt=f\rho-wp=-d\rho/dt.
\end{equation}
Using the normalization condition $p+\rho=1$, its steady state solution can be obtained as follows
\begin{equation}\label{eq2}
p=f/(f+w),\quad \rho=w/(f+w).
\end{equation}
Let $U_{eff}=up+f\rho$, $W_{eff}=wp+b\rho$, then the mean velocity of cargo can be obtained as follows
\begin{equation}\label{eq3}
V=(U_{eff}-W_{eff})l_0=[(u-w)p+(f-b)\rho]l_0=(uf-wb)l_0/(f+w),
\end{equation}
where $l_0$ is the step size of cargo.
The probabilities that cargo steps forward and backward are then
\begin{equation}\label{eq4}
\begin{aligned}
&p_+=\frac{U_{eff}}{U_{eff}+W_{eff}}=\frac{f(u+w)}{f(u+w)+w(f+b)},\cr
&p_-=1-p_+=\frac{w(f+b)}{f(u+w)+w(f+b)}.
\end{aligned}
\end{equation}

%On the other hand, since the direction of cargo motion depends on its previous one, one can easily show that the probabilities $p_+, p_-$ satisfies
%\begin{equation}\label{eq5}
%p_+=p_+p_{++}+p_-p_{-+},\quad p_-=p_+p_{+-}+p_-p_{--}.
%\end{equation}
%Where $p_{++}=u/(u+w)$ is the probability that cargo steps forward if it is in plus-state. Similarly $p_{-+}=f/(f+b)$, $p_{+-}=w/(u+w)$, and $p_{--}=b/(f+b)$.

Finally, the external load $F$ dependence of rate $u, w, f, b$ can be given by the following Bell approximation \cite{Bell1978,Fisher2001,Zhang20093,Zhang20112},
\begin{equation}\label{eq5}
u=u_0e^{-\epsilon_0 Fl_0/k_BT},\quad
w=w_0e^{(1-\epsilon_0)Fl_0/k_BT},\quad
f=f_0e^{-\epsilon_1 Fl_0/k_BT},\quad
b=b_0e^{(1-\epsilon_1) Fl_0/k_BT}.
\end{equation}
Where $\epsilon_{0}$ and $\epsilon_{1}$ are {\it load distribution factors} for the {\it plus-end} motor and {\it minus-end} motor, respectively. $k_B$ is Boltzmann constant, and $T$ is the absolute temperature.

\subsection*{Cargo Motion in one fixed optical trap}\label{IIB}
This special case is schematically depicted in Fig. \ref{FigSchemtic}(b). For convenience, the center of optical trap is assumed to be fixed at location $0$.
%Due to the difference of potential at different location, the step rates of motor protein depend on cargo location. Assuming the optical trap is centered at 0,
For this case, the potential of cargo depends on its location $n$. The potential difference between location $n$ and location $n+1$ is
$\Delta G_n=\kappa[(n+1)l_0]^2/2-\kappa(nl_0)^2/2=\kappa(n+1/2)l_0^2$. Similar as in \cite{Zhang20114}, at location $n$, the forward and backward step rates $u_n$ and $w_n$ of cargo in {\it plus-state}, as well as the step rates $f_n$ and  $b_n$ of cargo in {\it minus-state}, can be obtained as follows,
\begin{equation}\label{eq6}
u_n=ue^{-\epsilon_0 \Delta G_n/k_BT},\quad
w_n=we^{(1-\epsilon_0) \Delta G_{n-1}/k_BT},\quad
f_n=fe^{-\epsilon_1 \Delta G_n/k_BT},\quad
b_n=be^{(1-\epsilon_1) \Delta G_{n-1}/k_BT}.
\end{equation}
Where $u, w, f, b$ are cargo step rates when there is no optical trap and any other external load, which satisfy $u\gg w, b\gg f$.
For simplicity, this study assumes that $\epsilon_{0}$, $\epsilon_{1}$ are independent of cargo location $n$.

Let $p_n,\rho_n$ be the probabilities of finding cargo in {\it plus-state} $n^+$ and {\it minus-state} $n^-$, respectively. One can easily show $p_n,\rho_n$ are governed by the following equations
\begin{subequations}
\begin{equation}\label{eq7a}
dp_n/dt=u_{n-1}p_{n-1}+f_{n-1}\rho_{n-1}-(u_n+w_n)p_n,
\end{equation}
\begin{equation}\label{eq7b}
d\rho_n/dt=w_{n+1}p_{n+1}+b_{n+1}\rho_{n+1}-(f_n+b_n)\rho_n.
\end{equation}
\end{subequations}
The steady state solution of Eqs. (\ref{eq7a}, \ref{eq7b}) are as follows (for details see Sec. A of the supplemental materials)
\begin{subequations}
\begin{equation}\label{eqADD1a}
p_n=\left[\prod_{k=0}^{n-1}\left(\frac{(f_{k}+b_{k})u_{k}}{(u_{k+1}+w_{k+1})b_{k}}\right)\right]p_0,\quad
\textrm{for\ } n\ge1,
\end{equation}
\begin{equation}\label{eqADD1b}
p_n=\left[\prod_{k=n+1}^{0}\left(\frac{(u_{k}+w_{k})b_{k-1}}{(f_{k-1}+b_{k-1})u_{k-1}}\right)\right]p_0,\quad
\textrm{for\ } n\le-1,
\end{equation}
\begin{equation}\label{eqADD1c}
\rho_n=\frac{u_{n}}{b_{n}}p_{n}=\frac{u_{n}}{b_{n}}
\left[\prod_{k=0}^{n-1}\left(\frac{(f_{k}+b_{k})u_{k}}{(u_{k+1}+w_{k+1})b_{k}}\right)\right]p_0, \quad \textrm{for\ } n\ge1,
\end{equation}
\begin{equation}\label{eqADD1d}
\rho_n=\frac{u_{n}}{b_{n}}p_{n}=
\frac{u_{n}}{b_{n}}\left[\prod_{k=n+1}^{0}\left(\frac{(u_{k}+w_{k})b_{k-1}}{(f_{k-1}+b_{k-1})u_{k-1}}\right)\right]p_0, \quad \textrm{for\ } n\le-1,
\end{equation}
\begin{equation}\label{eqADD1e}
\rho_{0}=\frac{u_{0}}{b_{0}}p_{0}.
\end{equation}
\end{subequations}
Where $p_0$ can be obtained by the normalization condition $\sum_{n=-\infty}^{+\infty}(p_n+\rho_n)=1$.

The probability of finding cargo in {\it plus-state} is  $p=\sum_{n=-\infty}^{+\infty}p_n$, and the probability of finding cargo in {\it minus-state} is $\rho=\sum_{n=-\infty}^{+\infty}\rho_n$. The mean locations of cargo in {\it plus-state} and in {\it minus-state} are
\begin{subequations}
\begin{equation}\label{eq15a}
\langle n^+\rangle=\sum_{n=-\infty}^{+\infty}np_n/p,\quad
\langle n^-\rangle=\sum_{n=-\infty}^{+\infty}n\rho_n/\rho,
\end{equation}
\end{subequations}
respectively. The mean location of cargo is
\begin{equation}\label{eq15c}
\langle n\rangle=\sum_{n=-\infty}^{+\infty}n(p_n+\rho_n)
=p\langle n^+\rangle+\rho\langle n^-\rangle.
\end{equation}
Specially, for the {\it symmetric} cases $u=b, w=f$, i.e. the cargo is transported by two motors with the same step rates but different intrinsic directionality, one can verify that $\rho_n=p_{-n}$ and consequently $\rho=p, \langle n^-\rangle=-\langle n^+\rangle, \langle n\rangle=0$.

The external load dependence of rates $u_n, w_n, f_n, b_n$ [see Eq. (\ref{eq6})] means that, for a cargo towed by two motors in one fixed optical trap there are two critical values of the cargo location $n$,
\begin{equation}\label{eq34}
n_{c+}=\left\lceil \frac{k_BT}{\kappa l_0^2}\ln\frac{u}{w}+\frac12-\epsilon_0\right\rceil,\quad
n_{c-}=\left\lfloor \frac{k_BT}{\kappa l_0^2}\ln\frac{f}{b}+\frac12-\epsilon_1\right\rfloor,
\end{equation}
where $\lceil x\rceil$ is the smallest integer number which is not less than $x$, $\lfloor x\rfloor$ is the biggest integer number which is not bigger than $x$. The step rates of {\it plus-end} motor satisfy $u_n>w_n$ for $n<n_{c+}$, and $u_n\le w_n$ for $n\ge n_{c+}$. Similarly, the step rates of {\it minus-end} motor satisfy $b_n>f_n$ for $n>n_{c-}$, and $b_n\le f_n$ for $n\le n_{c-}$. The intrinsic directionality of {\it plus-end} motor ($u\gg w$) implies $n_{c+}>0$, and the intrinsic directionality of {\it minus-end} motor ($b\gg f$) implies $n_{c-}<0$.
%Rate differences $u_n-w_n$ and $b_n-f_n$ change their signs around $n_{c+}$ and $n_{c-}$, respectively.
%$\langle n\rangle$ is different with the trap center 0, and
Generally, the critical values $n_{c+}$ and $n_{c-}$ are different with the mean locations $\langle n^+\rangle$ and $\langle n^-\rangle$. %Usually $n_{c-}\le \langle n^-\rangle<0<\langle n^+\rangle<n_{c+}$.

In the following of this section, various mean first passage time (MFPT) problems about the cargo motion in fixed optical trap will be discussed.

\subsubsection*{Mean first passage time to one of the plus-state}\label{secIIB1}
Let $t_n^l$ and $\tau_n^l$ be MFPTs of cargo from plus-state $n^+$ and minus-state $n^-$ to plus-state $l^+$ respectively, then $t_n^l$ and $\tau_n^l$ satisfy \cite{Redner2001,Zhang20113}
\begin{subequations}
\begin{equation}\label{eq16a}
w_n\tau_{n-1}^l-(u_n+w_n)t_n^l+u_nt_{n+1}^l=-1,\quad \textrm{for\ } n\ne l,
\end{equation}
\begin{equation}\label{eq16b}
b_n\tau_{n-1}^l-(f_n+b_n)\tau_n^l+f_nt_{n+1}^l=-1,
\end{equation}
\end{subequations}
with one boundary condition $t_l^l=0$.

From Eq. (\ref{eq16a}) one can easily get
\begin{equation}\label{eq17}
\tau_{n-1}^l=\frac{u_n+w_n}{w_n}t_n^l-\frac{u_n}{w_n}t_{n+1}^l-\frac{1}{w_n}, \quad \textrm{for\ } n\ne l.
\end{equation}
Substituting (\ref{eq17}) into (\ref{eq16b}), one obtains
\begin{equation}\label{eq18}
b_n\left[\frac{u_n+w_n}{w_n}t_n^l-\frac{u_n}{w_n}t_{n+1}^l-\frac{1}{w_n}\right]
-(f_n+b_n)\left[\frac{u_{n+1}+w_{n+1}}{w_{n+1}}t_{n+1}^l-\frac{u_{n+1}}{w_{n+1}}t_{n+2}^l-\frac{1}{w_{n+1}}\right]
+f_nt_{n+1}^l=-1,
\end{equation}
i.e. %or, for the sake of convenience,
\begin{equation}\label{eq19}
B_nt_n^l-(B_n+F_n)t_{n+1}^l+F_nt_{n+2}^l=C_n,
\end{equation}
where
\begin{equation}\label{eq20}
B_n=\frac{(u_n+w_n)b_n}{w_n},\quad
F_n=\frac{(f_n+b_n)u_{n+1}}{w_{n+1}},\quad
C_n=\frac{b_n}{w_n}-\frac{f_n+b_n}{w_{n+1}}-1.
\end{equation}
Note, Eqs. (\ref{eq18}, \ref{eq19}) are established for $n\ne l-1, l$.

Meanwhile, from Eq. (\ref{eq16b}) one can get
\begin{equation}\label{eq21}
t_{n+1}^l=\frac{f_n+b_n}{f_n}\tau_n^l-\frac{b_n}{f_n}\tau_{n-1}^l-\frac{1}{f_n},
\end{equation}
and then by substituting Eq. (\ref{eq21}) into Eq. (\ref{eq16a}) one obtains
\begin{equation}\label{eq22}
w_n\tau_{n-1}^l-(u_n+w_n)
\left[\frac{f_{n-1}+b_{n-1}}{f_{n-1}}\tau_{n-1}^l-\frac{b_{n-1}}{f_{n-1}}\tau_{n-2}^l-\frac{1}{f_{n-1}}\right]
+u_n\left[\frac{f_n+b_n}{f_n}\tau_n^l-\frac{b_n}{f_n}\tau_{n-1}^l-\frac{1}{f_n}\right]
=-1,
\end{equation}
i.e.
\begin{equation}\label{eq23}
\hat B_n\tau_{n-2}^l-(\hat B_n+\hat F_n)\tau_{n-1}^l+\hat F_n\tau_{n}^l=\hat C_n,
\end{equation}
where
\begin{equation}\label{eq24}
\hat B_n=\frac{(u_n+w_n)b_{n-1}}{f_{n-1}},\quad
\hat F_n=\frac{(f_n+b_n)u_{n}}{f_{n}},\quad
\hat C_n=\frac{u_n}{f_n}-\frac{u_n+w_n}{f_{n-1}}-1.
\end{equation}
Eqs. (\ref{eq22}, \ref{eq23}) are established for $n\ne l$.

The procedure of getting MFPTs $t_n^l,\tau_n^l$ is as follows. {\bf (1) } Getting $t_n^l$ for $n\le l-1$ by Eq. (\ref{eq19}) and boundary condition $t_l^l=0$ (see  Sec. B of the supplemental materials). {\bf (2) } Getting $\tau_n^l$ for $n\le l-2$ by Eq. (\ref{eq17}). {\bf (3) } Getting $\tau_{l-1}^l$ from the special case of Eq. (\ref{eq16b}), i.e. $b_{l-1}\tau_{l-2}^l-(f_{l-1}+b_{l-1})\tau_{l-1}^l=-1$. {\bf (4) } Getting $\tau_n^l$ for $n\ge l$ by Eq. (\ref{eq23}) and boundary value $\tau_{l-1}^l$ obtained in {\bf (3) } (see Sec. C of the supplemental materials). {\bf (5) } Getting $t_n^l$ for $n\ge l+1$ by Eq. (\ref{eq21}).
This procedure can be summarized as follows
\begin{equation}\label{eqADD2}
\xLongrightarrow[t_l^l=0]{\textrm{Eq. (\ref{eq19})}}t_n^l (n\le l-1)
\xLongrightarrow{\textrm{Eq. (\ref{eq17})}}\tau_n^l (n\le l-2)
\xLongrightarrow[n=l-1]{\textrm{Eq. (\ref{eq16b})}}\tau_{l-1}^l
\xLongrightarrow{\textrm{Eq. (\ref{eq23})}}\tau_n^l (n\ge l)
\xLongrightarrow{\textrm{Eq. (\ref{eq21})}}t_n^l (n\ge l+1).
\end{equation}

\subsubsection*{Mean first passage time to one of the minus-state}\label{secIIB2}
Let $\bar t_n^l$ and $\bar\tau_{n}^l$ be the MFPTs of cargo from plus-state $n^+$ and minus-state $n^-$ to minus-state $l^-$, respectively. Similar as the discussion in Sec. \ref{secIIB1}, the MFPTs $\bar t_n^l$ and $\bar\tau_{n}^l$ satisfy the following equations
\begin{subequations}
\begin{equation}\label{eq25a}
w_n\bar \tau_{n-1}^l-(u_n+w_n)\bar t_n^l+u_n\bar t_{n+1}^l=-1,
\end{equation}
\begin{equation}\label{eq25b}
b_n\bar \tau_{n-1}^l-(f_n+b_n)\bar \tau_n^l+f_n\bar t_{n+1}^l=-1,\quad \textrm{for\ }n\ne l,
\end{equation}
\end{subequations}
with one boundary condition $\bar \tau_l^l=0$.
From Eq. (\ref{eq25a}) one can easily get
\begin{equation}\label{eq26}
\bar \tau_{n-1}^l=\frac{u_n+w_n}{w_n}\bar t_n^l-\frac{u_n}{w_n}\bar t_{n+1}^l-\frac{1}{w_n}.
\end{equation}
Substituting (\ref{eq26}) into (\ref{eq25b}), one obtains
\begin{equation}\label{eq27}
b_n\left[\frac{u_n+w_n}{w_n}\bar t_n^l-\frac{u_n}{w_n}\bar t_{n+1}^l-\frac{1}{w_n}\right]
-(f_n+b_n)\left[\frac{u_{n+1}+w_{n+1}}{w_{n+1}}\bar t_{n+1}^l-\frac{u_{n+1}}{w_{n+1}}\bar t_{n+2}^l-\frac{1}{w_{n+1}}\right]
+f_n\bar t_{n+1}^l=-1,
\end{equation}
i.e.
\begin{equation}\label{eq28}
B_n\bar t_n^l-(B_n+F_n)\bar t_{n+1}^l+F_n\bar t_{n+2}^l=C_n,
\end{equation}
%where
%\begin{equation}\label{eq29}
%B_n=\frac{(u_n+w_n)b_n}{w_n},\quad
%F_n=\frac{(f_n+b_n)u_{n+1}}{w_{n+1}},\quad
%C_n=\frac{b_n}{w_n}-\frac{f_n+b_n}{w_{n+1}}-1.
%\end{equation}
with $B_n, F_n, C_n$ given by  Eq. (\ref{eq20}). Note, Eqs. (\ref{eq27}, \ref{eq28}) are established for $n\ne l$.

Meanwhile, from Eq. (\ref{eq25b}) one can get
\begin{equation}\label{eq30}
\bar t_{n+1}^l=\frac{f_n+b_n}{f_n}\bar \tau_n^l-\frac{b_n}{f_n}\bar \tau_{n-1}^l-\frac{1}{f_n}, \quad \textrm{for\ }n\ne l,
\end{equation}
and then by substituting Eq. (\ref{eq30}) into Eq. (\ref{eq25a}) one obtains
\begin{equation}\label{eq31}
w_n\bar \tau_{n-1}^l-(u_n+w_n)
\left[\frac{f_{n-1}+b_{n-1}}{f_{n-1}}\bar \tau_{n-1}^l-\frac{b_{n-1}}{f_{n-1}}\bar \tau_{n-2}^l-\frac{1}{f_{n-1}}\right]
+u_n\left[\frac{f_n+b_n}{f_n}\bar \tau_n^l-\frac{b_n}{f_n}\bar \tau_{n-1}^l-\frac{1}{f_n}\right]
=-1,
\end{equation}
i.e.
\begin{equation}\label{eq32}
\hat B_n\bar \tau_{n-2}^l-(\hat B_n+\hat F_n)\bar \tau_{n-1}^l+\hat F_n\bar \tau_{n}^l=\hat C_n,
\end{equation}
%where
%\begin{equation}\label{eq33}
%\hat B_n=\frac{(u_n+w_n)b_{n-1}}{f_{n-1}},\quad
%\hat F_n=\frac{(f_n+b_n)u_{n}}{f_{n}},\quad
%\hat C_n=\frac{u_n}{f_n}-\frac{u_n+w_n}{f_{n-1}}-1.
%\end{equation}
with $\hat B_n, \hat F_n, \hat C_n$ given by Eq. (\ref{eq24}). Eqs. (\ref{eq31}, \ref{eq32}) are established for $n\ne l, l+1$.

The procedure of getting MFPTs $\bar t_n^l,\bar\tau_n^l$ is as follows. {\bf (1) } Getting $\bar \tau_n^l$ for $n\ge l+1$ by Eq. (\ref{eq32}) and boundary condition $\bar\tau_l^l=0$ (see Sec. D of the supplemental materials). {\bf (2) } Getting $\bar t_n^l$ for $n\ge l+2$ by Eq. (\ref{eq30}).  {\bf (3) } Getting $\bar t_{l+1}^l$ from the special case of Eq. (\ref{eq25a}), i.e. $-(u_{l+1}+w_{l+1})\bar t_{l+1}^l+u_{l+1}\bar t_{l+2}^l=-1$, {\bf (4) } Getting $\bar t_n^l$ for $n\le l$ by Eq. (\ref{eq28}) with boundary value $\bar t_{l+1}$ obtained in {\bf (3) } (see Sec. E of the supplemental materials). {\bf (5) } Getting $\bar\tau_n^l$ for $n\le l-1$ by Eq. (\ref{eq26}).
This procedure can be summarized as follows
\begin{equation}\label{eqADD2}
\xLongrightarrow[\bar\tau_l^l=0]{\textrm{Eq. (\ref{eq32})}}\bar \tau_n^l (n\ge l+1)
\xLongrightarrow{\textrm{Eq. (\ref{eq30})}}\bar t_n^l (n\ge l+2)
\xLongrightarrow[n=l+1]{\textrm{Eq. (\ref{eq25a})}}\bar t_{l+1}^l
\xLongrightarrow{\textrm{Eq. (\ref{eq28})}}\bar t_n^l (n\le l)
\xLongrightarrow{\textrm{Eq. (\ref{eq26})}}\bar\tau_n^l (n\le l-1).
\end{equation}

\subsubsection*{Mean first passage time to one given location}\label{secB3}
Let ${\cal T}_s^l$ be the MFPT of cargo from state $s$ to location $l$ (either {\it plus-state} $l^+$ or {\it minus-state} $l^-$), then one can easily show that
\begin{equation}\label{eq37}
{\cal T}_s^l=\left\{
\begin{array}{llll}
t_{k}^{l},\quad & \textrm{for\ } s=k^+ & \textrm{and}& k<l,\cr
\tau_{k}^{l},& \textrm{for\ } s=k^- & \textrm{and}& k<l,\cr
\bar t_{k}^{l},& \textrm{for\ } s=k^+ & \textrm{and}& k>l,\cr
\bar\tau_{k}^{l},& \textrm{for\ } s=k^- & \textrm{and}& k>l.
\end{array}\right.
\end{equation}
It is to say that if $k<l$, a cargo located at $k$ will first reach {\it plus-state} $l^+$ before reaching {\it minus-state} $l^-$. On the contrary, if $k>l$, it will first reach {\it minus-state} $l^-$. Finally, the mean oscillation period $T$ of cargo in fixed optical trap can be approximated as follows
\begin{equation}\label{eqADD3}
T\approx\tau_{0}^{0}+\bar t_{0}^{0},
%T\approx\tau_{0^-}^{0^+}+\bar t_{0^+}^{0^-},
\end{equation}
see Sec. F of the supplemental materials for its expression.

\section*{Results}\label{result}

\begin{figure}
\begin{center}
  \includegraphics[width=450pt]{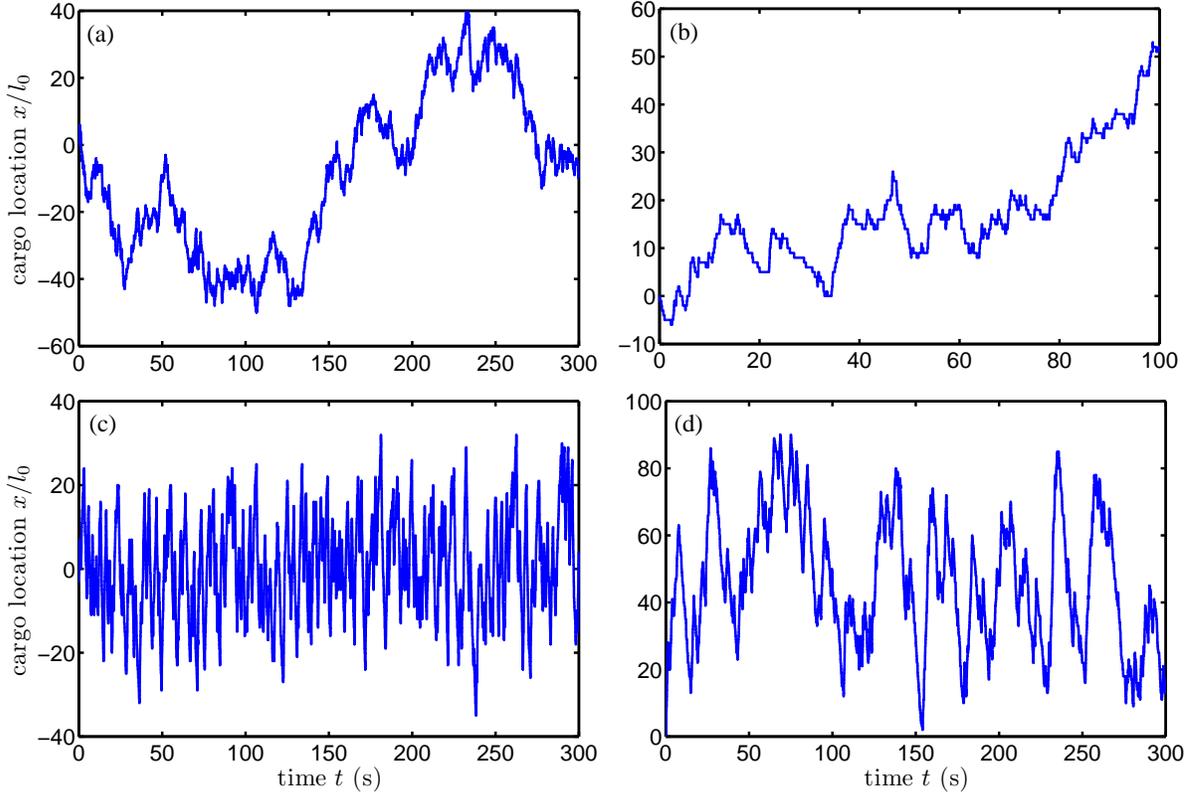}\\
\end{center}
  \caption{Trajectory samples of cargo motion by two motors under constant load (a, b), and in one fixed optical trap (c, d). For the symmetric cases (where the step rates of the plus motor are the same as the ones of the minus motor, i.e. $u=b$, $w=f$), the cargo will oscillate around its initial location (a). While for the asymmetric cases, the cargo will have nonzero mean velocity (b). If the cargo is put in one fixed optical trap and transported by two {\it symmetric} motors, it will   oscillate around the trap center (c). But for the asymmetric cases, the oscillation center may be different from the trap center. For parameter values used in the simulations see Tab. I. }\label{Fig2}
\end{figure}

\begin{figure}
\begin{center}
  \includegraphics[width=450pt]{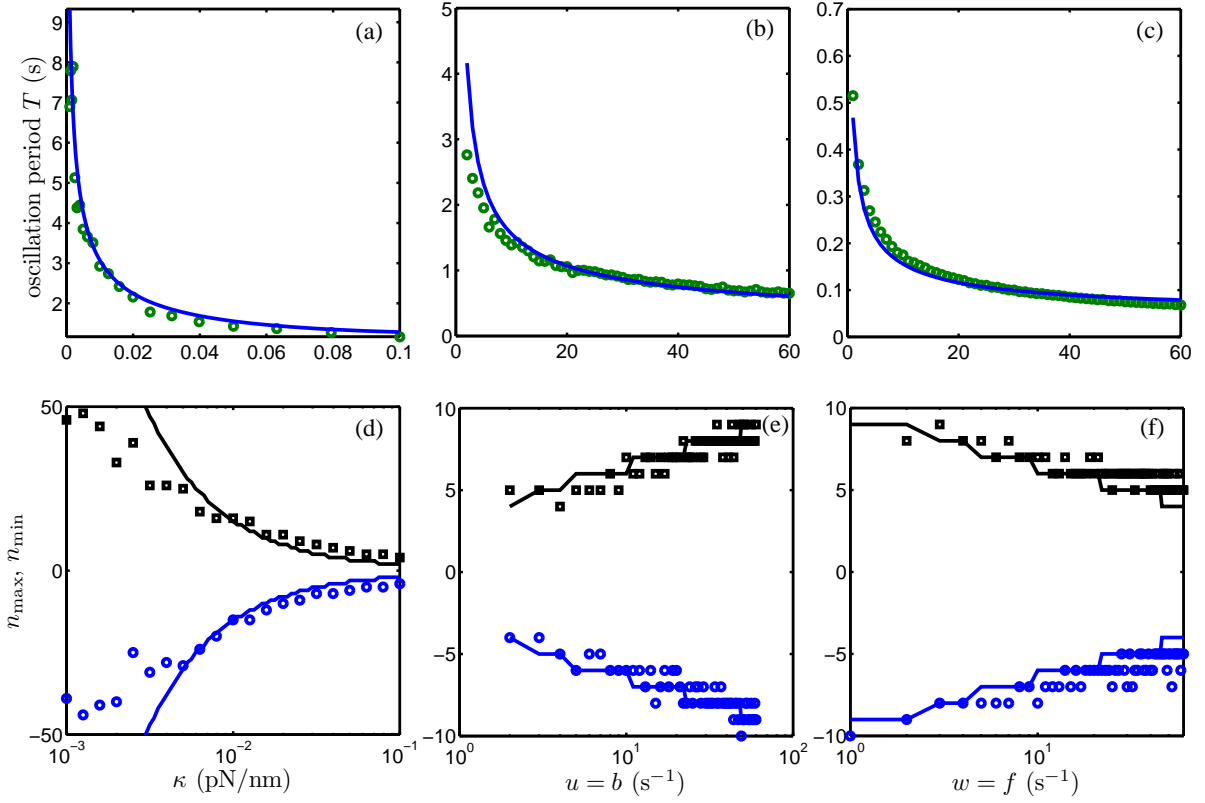}\\
\end{center}
  \caption{In fixed optical trap, the mean oscillation period $T$ of cargo decreases with trap stiffness $\kappa$, forward rates $u=b$, and backward rates $w=f$ (in fact, $\log T$ decreases almost linearly with $\log\kappa$, $\log u=\log b$, and $\log w=\log f$). The oscillation amplitude $n_{\max}-n_{\min}$ decreases with stiffness $\kappa$ and backward rates $w=f$, but increases with forward rates $u=b$. Here $n_{\max}$ and $n_{\min}$ are the $\max$ and $\min$ locations that cargo can reaches. The circles and squares are obtained by Monte Carlo simulations. In (a, b, c), the solid curves are obtained by formulation (\ref{eqADD3}). The solid lines in (d) are obtained by $n_{c+}, n_{c-}$ given in Eq. (\ref{eq34}), and the solid lines in (e, f) are obtained by $n_{c+}+3, n_{c-}-3$, respectively. For parameter values see Tab. I. }\label{Fig3}
\end{figure}

%\subsection{Monte Carlo simulation}
For cargo motion under no external load, Monte Carlo simulations show that, if the cargo is transported by two {\it symmetric} motors, i.e., the {\it plus-end} motor and the {\it minus-end} motor have the same step rates, $u=b, w=f$, the cargo will oscillate [Fig. \ref{Fig2}(a)]. While for the {\it asymmetric} cases, the cargo has non-zero mean velocity [see Fig. \ref{Fig2}(b)]. On the other hand, if the cargo is put into one fixed optical trap, and transported by two {\it symmetric} motors, it will oscillate around the trap center with relatively high frequency [Fig. \ref{Fig2}(c)]. Meanwhile, if the trapped cargo is transported by two {\it asymmetric} motors, it will also oscillate but its oscillation center may be different with the trap center [Fig. \ref{Fig2}(d)]. Both Monte Carlo simulations and theoretical calculations show that, for a cargo transported by two {\it symmetric} motors and put in one optical trap, its oscillation period $T$ decreases with trap stiffness $\kappa$, motor forward step rates $u=b$, and motor backward step rates $w=f$ [Fig. \ref{Fig3}(a-c)]. Its oscillation amplitude increases with the motor forward step rates $u=b$, but decreases with both the motor backward step rates $u=b$ and the trap stiffness $\kappa$, since high backward step rates and high trap stiffness will prohibit the cargo from moving too far from the trap center [Fig. \ref{Fig3}(d-f)].

Let
\begin{equation}\label{eqnew32}
p=\sum_{n=-\infty}^{\infty} p_n,\quad
\rho=\sum_{n=-\infty}^{\infty} \rho_n,\quad
P_+=\sum_{n>0} (p_n+\rho_n),\quad
P_-=\sum_{n<0} (p_n+\rho_n).
\end{equation}
Then $p$ is the probability of finding cargo in {\it plus-state}, $P_+$ is the probability that cargo location $n>0$ (the center of optical trap is assumed to be at location 0). The meanings of $\rho$ and $P_-$ are similar. Both Monte Carlo simulations and theoretical calculations show that, for a cargo transported by two {\it symmetric} motors, the ratios $p/\rho$ and $P_+/P_-$ are always one, and they do not change with trap stiffness $\kappa$, forward step rates $u=b$, and backward step rates $w=f$ [Fig. S1].

Our results also show that, for cargo motion in optical trap by two {\it asymmetric} motors, its oscillation period $T$ decreases with trap stiffness $\kappa$ and forward step rate $u$, but may not change monotonically with backward step rate $w$ [Figs. S2(a), S3(a), S4(a)]. But similar as the {\it symmetric} cases, cargo oscillation amplitude of the {\it asymmetric} cases decreases with trap stiffness $\kappa$ and backward step rate $w$, and increases with the forward step rate $u$ [Figs. S2(d), S3(d), S4(d)]. The results in Figs. S3(d),\ and S4(d) imply that, the maximal location $n_{\max}$ that cargo might reach toward the {\it plus-end} of microtubule depends only on the step rates $u, w$ of the {\it plus-end} motor, and similarly the minimal location $n_{\min}$ that cargo might reach towards the {\it minus-end} of the microtubule depends only on the step rates $b, f$ of the {\it minus-end} motor. From the results given in Figs. S2(b,c), S3(b,c), and S4(b,c) one can also see that, different from the {\it symmetric} cases given in Fig. S1, both the ratio $p/\rho$ and ratio $P_+/P_-$ depend on trap stiffness $\kappa$, forward step rate $u$, and backward step rate $w$.

\begin{figure}
\begin{center}
  \includegraphics[width=450pt]{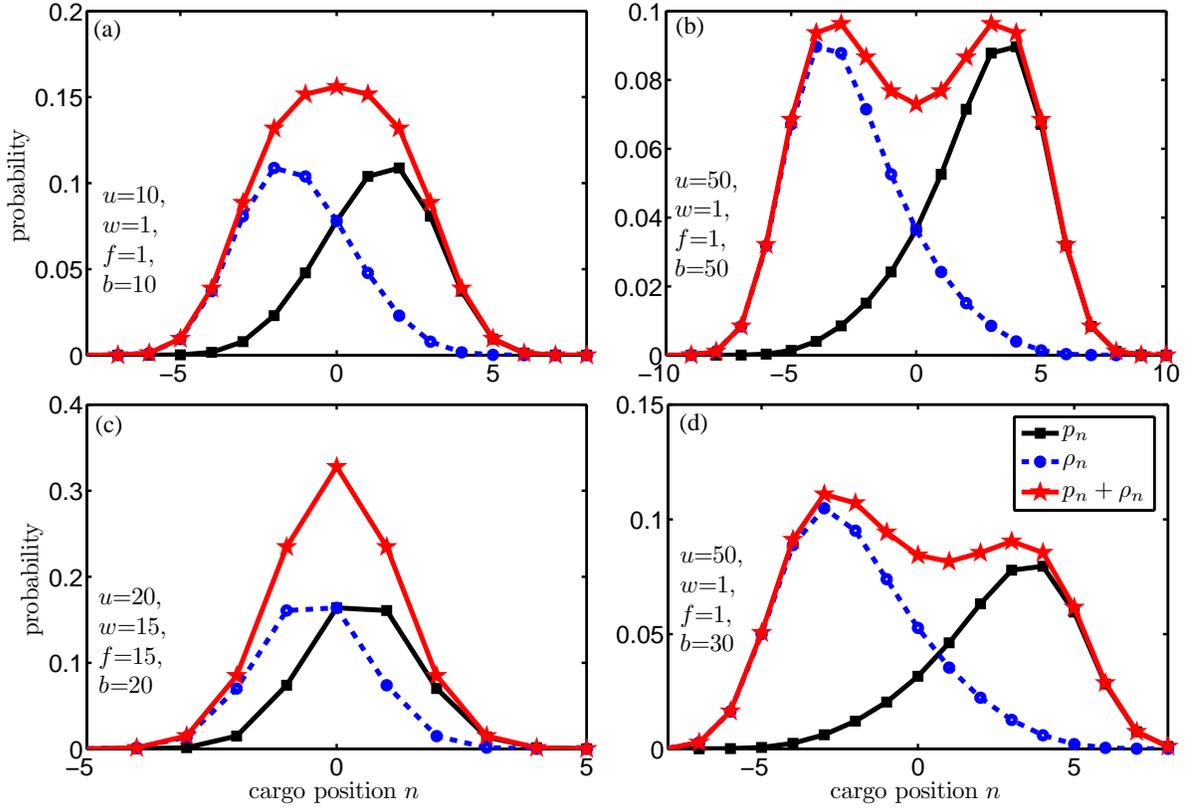}\\
\end{center}
  \caption{Samples of probability $p_n$ and $\rho_n$ for finding cargo in {\it plus-state} and {\it minus-state}. For the symmetric cases probabilities $p_n$ and $\rho_n$ are mirror symmetry to each other (a, b, c). Their sum $p_n+\rho_n$, the probability of finding cargo at location $n$, might has one maximum [at the center of optical trap, see (a, c)] or two symmetric maximum [see (b)]. (d) is one sample for the asymmetric cases. For parameter values see Tab. I. }\label{Fig4}
\end{figure}

To show more details about the dependence of cargo oscillation on trap stiffness $\kappa$ and motor step rates, examples of probabilities $p_n, \rho$, and their summation $p_n+\rho_n$ are plotted in Fig. \ref{Fig4} and Fig. S5. For either {\it symmetric} cases or {\it asymmetric} cases, the probability profiles are flat for low trap stiffness $\kappa$, indicating that the cargo can reach a farther location from the oscillation center (i.e., with large oscillation amplitude)[Fig. S5]. Similar changes can also be found with the increase of motor forward step rates $u$ or $f$ [Fig. \ref{Fig4}(a, b, d)]. Meanwhile, with the increase of motor backward step rates $w$ or $f$, the probability profile will become more sharp [Fig. \ref{Fig4}(c)]. For the {\it asymmetric} cases, the most likely location of cargo may be different from the trap center [Fig. S5(c)]. One interesting phenomenon displayed in Fig. \ref{Fig4}(b, d) is that, for either the {\it symmetric} cases or the {\it asymmetric} cases, when motor forward step rates $u, b$ are high, the summation of probability $p_n+\rho_n$ may has two local maxima, indicating that cargo motion in the positive location ($n>0$) is mainly dominated by the plus motor, while its motion in the negative location ($n<0$) is mainly dominated by the minus motor.

\begin{figure}
\begin{center}
  \includegraphics[width=450pt]{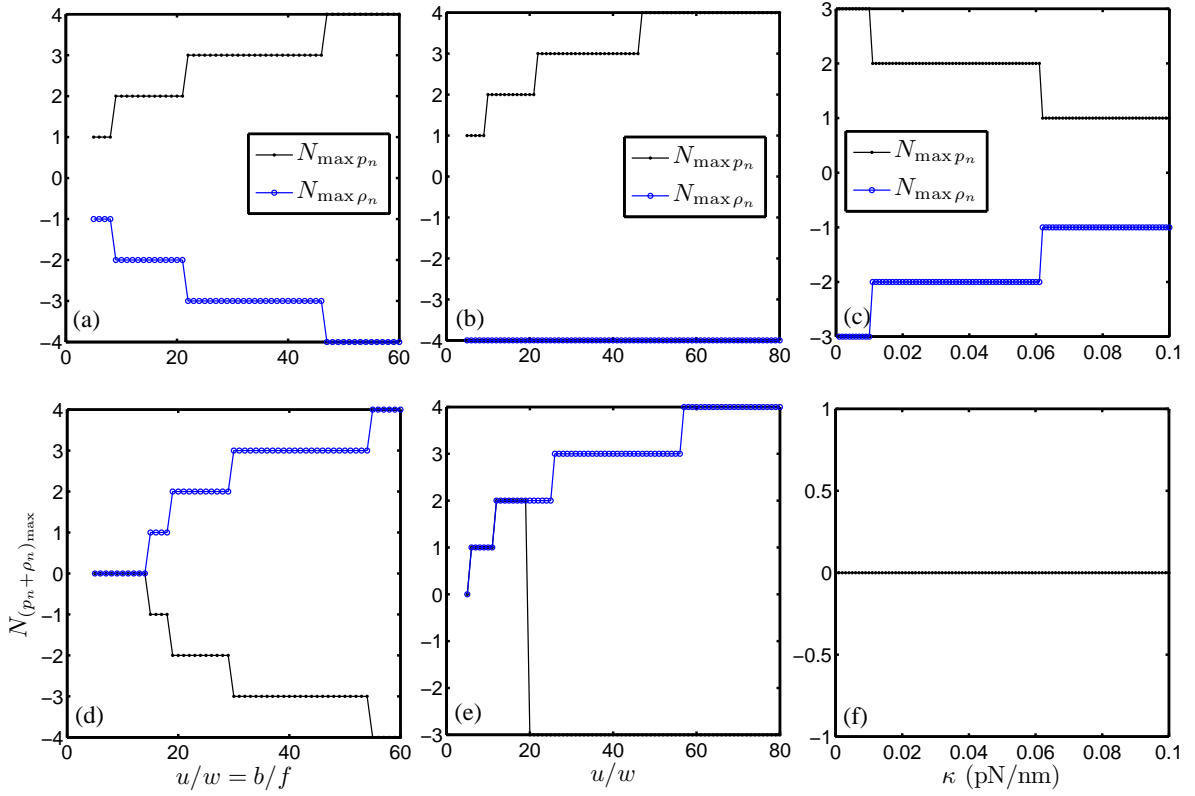}\\
\end{center}
  \caption{The location $N_{\max p_n}, N_{\max \rho_n}, N_{{(p_n+\rho_n)}_{\max}}$ that probabilities $p_n$, $\rho_n$ and their summation $p_n+\rho_n$ reach their maximum. With the increase of rate ratio $u/w=b/f$ both $N_{\max p_n}$ and $N_{\max \rho_n}$ leave far away from the trap center (a). (b) implies that $N_{\max p_n}$ increases with ratio $u/w$, but $N_{\max \rho_n}$ is independent of it. With the increase of trap stiffness $\kappa$, both $N_{\max p_n}$ and $N_{\max \rho_n}$ come close the the trap center (c). (d, e) show that, with the increase of rate ratio $u/w=b/f$ or rate ratio $u/w$ only, the number of maximum of probability $p_n+\rho_n$ of finding cargo at location $n$ may change. But (f) implies that $N_{{(p_n+\rho_n)}_{\max}}$ is independent of trap stiffness $\kappa$.  For parameter values see Tab. I. }\label{Fig5}
\end{figure}

Let $N_{\max p_n}, N_{\max \rho_n}, N_{{(p_n+\rho_n)}_{\max}}$ be the locations at which probabilities $p_n$, $\rho_n$ and their summation $p_n+\rho_n$ reach their maxima, respectively. The results plotted in Fig. \ref{Fig5}(a) show that, for {\it symmetric} motion, $N_{\max \rho_n}=-N_{\max p_n}$ and their absolute values increase with the forward to backward step rate ratio $u/w=b/f$. The results in Fig. \ref{Fig5}(d) show that, for low step rate ratio $u/w=b/f$, the total probability $p_n+\rho_n$ has only one maximum which lies at the trap center. However, with increase of these ratios, $N_{{(p_n+\rho_n)}_{\max}}$ has one {\it symmetric} bifurcation, and its absolute value %of $N_{{(p_n+\rho_n)}_{\max}}$ %(i.e. the locations of maxima of probability $p_n+\rho_n$,
(see Fig. \ref{Fig4}) increases with these step ratios. For {\it asymmetric} case [see Fig. \ref{Fig5}(b)], $N_{\max p_n}$ increases with step rate ratio $u/w$, but $N_{\max \rho_n}$ is independent of it. Which means that, similar as the properties of $n_{\max}$ and $n_{\min}$ displayed in Figs. S3 and S4, $N_{\max p_n}$ depends only on step rates of the {\it plus-end} motor, and $N_{\max \rho_n}$ depends only on step rates of the {\it minus-end} motor. For {\it asymmetric} cases, with the increase of rate ratio $u/w$,  $N_{{(p_n+\rho_n)}_{\max}}$ has also one bifurcation, see Fig. \ref{Fig5}(e). But one of the two values (the negative one) does not change with $u/w$. Which means that, the negative one of $N_{{(p_n+\rho_n)}_{\max}}$ depends only on properties of the {\it minus-end} motor. Similarly, the positive one of $N_{{(p_n+\rho_n)}_{\max}}$ depends only on properties of the {\it plus-end} motor. So both the properties of amplitude $n_{\max}, n_{\min}$ and the most likely locations $N_{\max p_n}, N_{\max \rho_n}, N_{{(p_n+\rho_n)}_{\max}}$ %of probabilities $p_n, \rho_n, p_n+\rho_n$
indicate that, the {\it plus-end} directed motion of cargo is mainly determined by the {\it plus-end} motor, and the {\it minus-end} directed motion is mainly determined by the {\it minus-end} motor, which is one of the main differences with other tug-of-war models \cite{Lipowsky2008,Kunwar2010,Driver2010,Zhang20114}, and this result is consistent with the experimental phenomena \cite{Gennerich2006,Soppina2009,Leidel2012}.
Finally, the results in Fig. \ref{Fig5}(c) show that, the absolute values of $N_{\max p_n}, N_{\max \rho_n}$ decrease with trap stiffness $\kappa$, and Fig. \ref{Fig5}(f) shows $N_{{(p_n+\rho_n)}_{\max}}$ does not change with stiffness $\kappa$. So trap stiffness can change the oscillation amplitude and the oscillation period (see Figs. \ref{Fig3}, S2, and S5), but will not change the most likely location $N_{{(p_n+\rho_n)}_{\max}}$ of the cargo. Further calculations of probabilities $p, \rho$ show that, for the {\it symmetric} cases both $p_{\max}=\rho_{\min}$ and $(p+\rho)_{\min}$ decrease with step rate ratio $u/w=b/f$, and increase with trap stiffness $\kappa$ [see Figs. S6(a,d)]. Since with large rate ratio $u/w=b/f$ and small stiffness $\kappa$, the cargo will oscillate with large amplitude. For the {\it asymmetric} cases, $p_{\max}\ne\rho_{\min}$, $p_{\max}$ decreases but $\rho_{\min}$ increases with the step rate ratio $u/w$ (i.e. with the increase of the directionality of the {\it plus-end} motor). Since with large rate ratio $u/w$, the {\it plus-end} motor has high directionality, and so the cargo moves fast in the {\it plus-state}, which means that the probability $p_n$ will be flat with large $u/w$. The plots in Fig. S6(c) show that, although the total probability $p_n+\rho_n$ has two maxima, with the change of rate ratio $u/w$, the most likely location of cargo may change from one side of the trap center to another side.

\begin{figure}
\begin{center}
  \includegraphics[width=450pt]{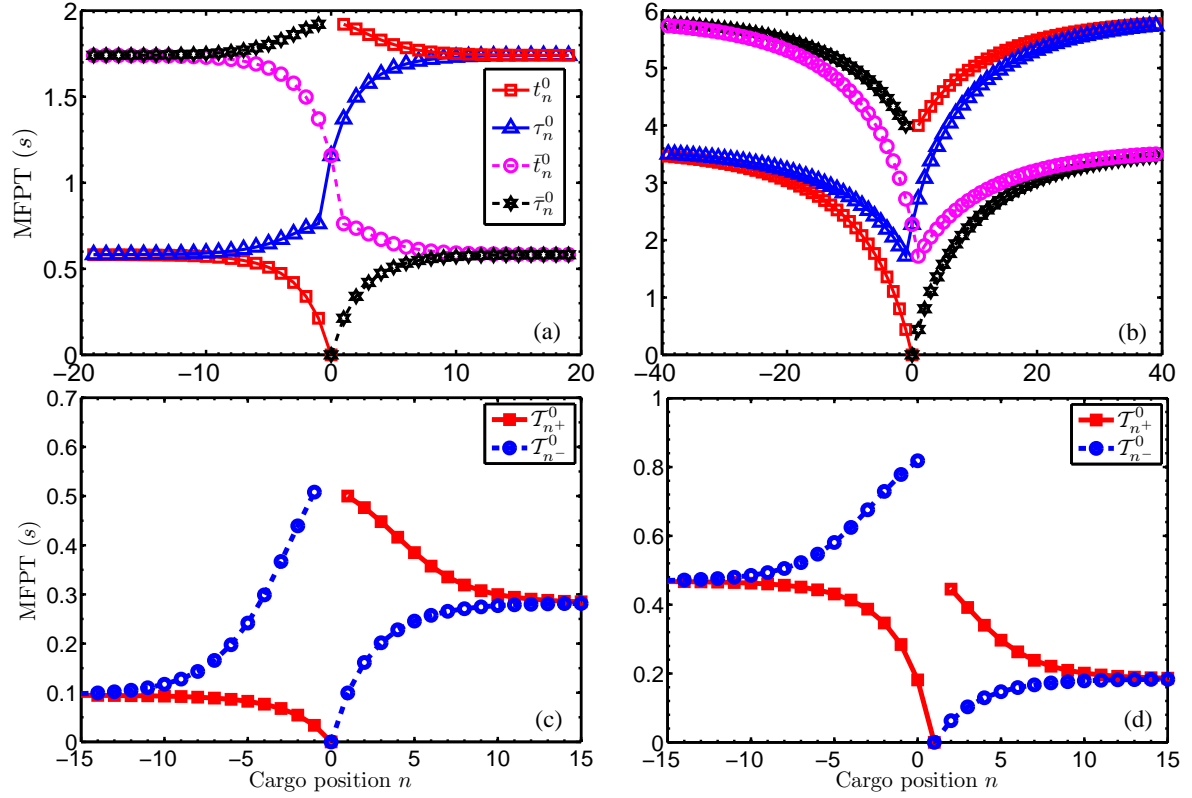}\\
\end{center}
  \caption{Samples of MFPTs $t_n^0, \tau_n^0$ to {\it plus-state} $0^+$, MFPTs $\bar t_n^0, \bar\tau_n^0$ to {\it minus-state} $0^-$ (a, b), and MFPT ${\cal T}_l^0$ from state $l$ to location 0 (c, d). For high trap stiffness $\kappa$,  $t_{n<0}^0<\tau_{m<0}^0<\tau_{l\ge0}^0<t_{k>0}^0$ for MPFTs to {\it plus-state} $0^+$, and symmetric relations hold for MFPTs  to {\it minus-state} $0^-$, see (a). But for low trap stiffness, all MFPTs $t_n^0, \tau_n^0, \bar t_n^0, \bar\tau_n^0$ increases with the distance between $n$ and trap center 0, see (b). Which means that, for different trap stiffness $\kappa$, the trajectories of cargo from state $n^+$ or $n^-$ to state $0^+$ or $0^-$  are different.  (c, d) are MFPTs for one cargo (transported by two {\it asymmetric} motors) from state $n^+$ or $n^-$ to location 0 ({\it plus-state} $0^+$ or $0^-$) and location 1 ({\it plus-state} $1^+$ or $1^-$).  The MFPT ${\cal T}_{n}^0$ is obtained by formulation (\ref{eq37}).  For parameter values see Tab. I. }\label{Fig6}
\end{figure}

Finally, several examples of MFPTs $t_n^l, \tau_n^l, \bar t_n^l, \bar\tau_n^l$ are plotted in Fig. \ref{Fig6}(a,b) and Figs. S7, S8(a,b), S9-S12, and examples of MFPTs ${\cal T}_{n^\pm}^l$ are plotted in Fig. \ref{Fig6}(c,d) and Fig. S8(c,d). If $m<n<l$, then $t_n^l\le t_m^l, \tau_m^l\le \tau_n^l$, $\bar t_n^l\le\bar t_m^l, \bar\tau_m^l\le\bar\tau_n^l$, and ${\cal T}_{n^+}^l\le{\cal T}_{m^+}^l, {\cal T}_{m^-}^l\le{\cal T}_{n^-}^l$.
If $l<n<m$, then $t_n^l\ge t_m^l, \tau_m^l\ge \tau_n^l$, $\bar t_n^l\ge\bar t_m^l, \bar\tau_m^l\ge\bar\tau_n^l$, and ${\cal T}_{n^+}^l\ge{\cal T}_{m^+}^l, {\cal T}_{m^-}^l\ge{\cal T}_{n^-}^l$.
Moreover, if the trap stiffness $\kappa$ is high and the motor step rate ratios $u/w$ and $b/f$ are large, then $t_m^l\le\tau_m^l, \bar t_m^l\le\bar\tau_m^l$, ${\cal T}_{m^+}^l\le{\cal T}_{m^-}^l$ for $m<n<l$, and $t_m^l\ge\tau_m^l, \bar t_m^l\ge\bar\tau_m^l$, ${\cal T}_{m^+}^l\ge{\cal T}_{m^-}^l$ for $l<n<m$, see Fig. \ref{Fig6}(a,c,d) and Figs. S7(a,b), S8(c,d),S9, S10(a), S11(b,c,d), S12(a). %Which means that, for large trap stiffness $\kappa$, and rate ratios $u/w, b/f$, the cargo will oscillate along one steady state cycle, the fluctuation in cargo oscillation is small.

\section*{Concluding Remarks}\label{conclusion}
Recent experimental observations by Leidel {\it et al.} \cite{Leidel2012} show that, in living cells cargo moves along microtubule with memory, i.e., its motion direction depends on its previous motion trajectory. In this study, such cargo transportation is theoretically studied by assuming that the cargo has the least memory, i.e. its motion direction depends only on its behavior in its last step. The cargo will be more likely to step forward/backward if it came to its present location by one forward/backward step. Two cases are mainly discussed: {\bf (I)} cargo moves under constant load, and {\bf (II)} cargo moves in one fixed optical trap. For each cases, two kinds of motion are addressed: {\bf (i) } {\it symmetric} motion, in which cargo is transported by two species of motor protein which have the same forward/backward step rates but with different intrinsic directionality, {\bf (ii)} {\it asymmetric} motion, in which cargo is transported by two species of motor protein with different forward/backward step rates. For the {\it symmetric} motion {\bf (i) } of case {\bf (I)}, the mean velocity of cargo is zero. But, due to the existence of memory, cargo can move unidirectionally for a large distance before switching its direction. One can easily understand that, for the {\it asymmetric} motion {\bf (ii) } of {\bf (I)}, the directionality of cargo with memory is better than that in the usual tug-of-war model by two different motor species \cite{Lipowsky2008, Driver2010,Zhang20114}. For the motion in one fixed optical trap, i.e. case {\bf (II)}, cargo will oscillate. For the {\it symmetric} motion {\bf (i)}, the oscillation center is the same as the trap center, but for the {\it asymmetric} motion {\bf (ii) }, this oscillation center is generally different from the trap center. Usually the oscillation period decreases with the trap stiffness $\kappa$ and motor step rates. Meanwhile, the oscillation amplitude decreases with trap stiffness $\kappa$ and motor backward step rates $w, f$, but increases with motor forward step rates $u, b$. The probability $p_n+\rho_n$ of finding cargo at location $n$ may have only one maximum, which is the same as the trap center for {\it symmetric} motion {\bf (i)} but different with the trap center for {\it asymmetric} motion {\bf (ii)}. Meanwhile, the probability $p_n+\rho_n$ may also have two maxima. For {\it symmetric} motion {\bf (i)}, these two maxima are located symmetrically on the two side of the trap center, and their corresponding values of probability $p_n+\rho_n$ are the same. However, for the {\it asymmetric} motion {\bf (ii)}, these two maxima are generally not symmetrically located around the trap center, and their corresponding probabilities may be greatly different. With the change of ratio of motor forward to backward step rates, the maximum with the larger value of probability $p_n+\rho_n$ may transfer from one side of the trap center to another side. This study will be helpful to understand the high directionality of cargo motion in living cells by cooperation of two species of motor protein. Meanwhile, more generalized model can also be employed to discuss this cargo transportation process, in which the cargo is assumed to have long memory, its forward and backward step rates depend on how long it has kept moving in its present direction.

\section*{Acknowledgments}
%\begin{acknowledgements}
This study was supported by the Natural Science Foundation of China (Grant No. 11271083), Natural Science Foundation of Shanghai (Grant No. 11ZR1403700), and the National Basic Research Program of China (National \lq\lq973" program, project No. 2011CBA00804).
%\end{acknowledgements}

\newpage

%\newpage

\section*{Tables}

\setcounter{table}{0}
\renewcommand{\thetable}{\Roman{table}}

\begin{table}[!ht]
\begin{center}
\caption{The values of rates $u, w, f, b$ (in unit s$^{-1}$) and optical trap stiffness $\kappa$ (pN/nm) used in the plots of Figs. \ref{Fig2}-\ref{Fig6}. The symbol $\ast$ means that the corresponding parameter is not used in the plot, and symbol $\checkmark$ means this parameter is one variable in the corresponding plot. Other parameters used in the plots are $\epsilon_0=\epsilon_1=0.5$, $l_0=8$ nm, and $k_BT=4.12$ pN$\cdot$nm. The stiffness $\kappa$ of the trap used in recent experiment of Leidel {\it el al.} is around $0.02-0.09$ pN/nm \cite{Leidel2012}.}
\begin{tabular}{l|ccccr}
%  \hline
  \hline\hline
   & $u$ & $w$\ &\ $f$\ &\ $b$\ &\qquad $\kappa$\ \ \\
   \hline
  Fig. \ref{Fig2}(a) & 5 & 2 & 2 & 5 & $\ast$ \\
  Fig. \ref{Fig2}(b) & 5 & 2 & 1 & 2 & $\ast$ \\
  Fig. \ref{Fig2}(c) & 20 & 1 & 1 & 20 & 0.004\\
  Fig. \ref{Fig2}(d) & 20 & 1 & 1 & 5 & 0.001\\
  \hline
  Fig. \ref{Fig3}(a,d) & 10 & 1 & 1 & 10 & $\checkmark$ \\
  Fig. \ref{Fig3}(b,e) &$\checkmark$ & 1 & 1 &$\checkmark$ & 0.05\\
  Fig. \ref{Fig4}(c,f) & 100 &$\checkmark$ &$\checkmark$ & 100 & 0.05\\
  \hline
  Fig. \ref{Fig4}(a) & 10 & 1 & 1 & 10 & 0.05 \\
  Fig. \ref{Fig4}(b) & 50 & 1 & 1 & 50 & 0.05 \\
\end{tabular}
\qquad\qquad
\begin{tabular}{l|ccccr}
%  \hline
  \multicolumn{6}{c}{ }\\
  Fig. \ref{Fig4}(c) & 20 & 15 & 15 & 20 & 0.05\\
  Fig. \ref{Fig4}(d) & 50 & 1 & 1 & 30 & 0.05\\
  \hline
  Fig. \ref{Fig5}(a,d) &$\checkmark$ & 1 & 1 &$\checkmark$ & 0.05 \\
  Fig. \ref{Fig5}(b,e) &$\checkmark$ & 1 & 1 & 50& 0.05\\
  Fig. \ref{Fig5}(c,f) & 10& 1 & 1 & 10&$\checkmark$\\
  \hline
  Fig. \ref{Fig6}(a) & 5 & 1 & 1 & 5 & 0.05 \\
  Fig. \ref{Fig6}(b) & 5 & 1 & 1 & 5 & 0.01 \\
  Fig. \ref{Fig6}(c) & 30 & 1 & 1 & 10 & 0.05\\
  Fig. \ref{Fig6}(d) & 10 & 1 & 1 & 10 & 0.05\\
  \hline
  \hline
\end{tabular}%\label{table2}
\end{center}
%\caption{The values of rates $u, w, f, b$ (in unit s$^{-1}$) and optical trap stiffness $\kappa$ (pN/nm) used in the plots of Figs. \ref{Fig2}-\ref{Fig6}. The symbol $\ast$ means that the corresponding parameter is not used in the plot, and symbol $\checkmark$ means this parameter is one variable in the corresponding plot. Other parameters used in the plots are $\epsilon_0=\epsilon_1=0.5$, $l_0=8$ nm, and $k_BT=4.12$ pN$\cdot$nm. The stiffness $\kappa$ of the trap used in recent experiment of Leidel {\it el al.} is around $0.02-0.09$ pN/nm \cite{Leidel2012}.}
\end{table}\label{table1}

%\newpage
%
%\section*{Figures}

\begin{thebibliography}{10}
\providecommand{\url}[1]{\texttt{#1}}
\providecommand{\urlprefix}{URL }
\expandafter\ifx\csname urlstyle\endcsname\relax
  \providecommand{\doi}[1]{doi:\discretionary{}{}{}#1}\else
  \providecommand{\doi}{doi:\discretionary{}{}{}\begingroup
  \urlstyle{rm}\Url}\fi
\providecommand{\bibAnnoteFile}[1]{%
  \IfFileExists{#1}{\begin{quotation}\noindent\textsc{Key:} #1\\
  \textsc{Annotation:}\ \input{#1}\end{quotation}}{}}
\providecommand{\bibAnnote}[2]{%
  \begin{quotation}\noindent\textsc{Key:} #1\\
  \textsc{Annotation:}\ #2\end{quotation}}
\providecommand{\eprint}[2][]{\url{#2}}

\bibitem{Bray2001}
Bray D (2001) Cell movements: from molecules to motility, 2nd Edn.
\newblock Garland, New York.
\input{Bray2001}

\bibitem{Howard2001}
Howard J (2001) Mechanics of Motor Proteins and the Cytoskeleton.
\newblock Sinauer Associates and Sunderland, MA.
\input{Howard2001}

\bibitem{Block1990}
Block SM, Goldstein LSB, Schnapp BJ (1990) Bead movement by single kinesin
  molecules studied with optical tweezers.
\newblock Nature 348: 348-352.
\input{Block1990}

\bibitem{Vale2003}
Vale RD (2003) The molecular motor toolbox for intracellular transport.
\newblock Cell 112: 467-480.
\input{Vale2003}

\bibitem{Mallik2004}
Mallik R, Carter BC, Lex SA, King SJ, Gross SP (2004) Cytoplasmic dynein
  functions as a gear in response to load.
\newblock Nature 427: 649-652.
\input{Mallik2004}

\bibitem{Hua1997}
Hua W, Young EC, Fleming ML, Gelles J (1997) Coupling of kinesin steps to {ATP}
  hydrolysis.
\newblock Nature 388: 390-393.
\input{Hua1997}

\bibitem{Schnitzer1997}
Schnitzer MJ, Block SM (1997) Kinesin hydrolyses one {ATP} per 8-nm step.
\newblock Nature 388: 386-390.
\input{Schnitzer1997}

\bibitem{Coy1999}
Coy DL, Wagenbach M, Howard J (1999) Kinesin takes one 8-nm step for each {ATP}
  that it hydrolyzes.
\newblock J Biol Chem 274: 3667-3671.
\input{Coy1999}

\bibitem{Gennerich2007}
Gennerich A, Carter AP, Reck-Peterson SL, Vale RD (2007) Force-induced
  bidirectional stepping of cytoplasmic dynein.
\newblock Cell 131: 952-965.
\input{Gennerich2007}

\bibitem{Asbury2003}
Asbury CL, Fehr AN, Block SM (2003) Kinesin moves by an asymmetric
  hand-over-hand mechanism.
\newblock Science 302: 2130-2134.
\input{Asbury2003}

\bibitem{Toba2006}
Toba S, Watanabe TM, Yamaguchi-Okimoto L, Toyoshima YY, Higuchi H (2006)
  Overlapping hand-over-hand mechanism of single molecular motility of
  cytoplasmic dynein.
\newblock Proc Natl Acad Sci USA 103: 5741-5745.
\input{Toba2006}

\bibitem{Guydosh2009}
Guydosh NR, Block SM (2009) Direct observation of the binding state of the
  kinesin head to the microtubule.
\newblock Nature 461: 125-128.
\input{Guydosh2009}

\bibitem{Lipowsky2005}
Klumpp S, Lipowsky R (2005) Cooperative cargo transport by several molecular
  motors.
\newblock Proc Natl Acad Sci USA 102: 17284-17289.
\input{Lipowsky2005}

\bibitem{Lipowsky2008}
M\"{u}ller MJI, Klumpp S, Lipowsky R (2008) Tug-of-war as a cooperative
  mechanism for bidirectional cargo transport by molecular motors.
\newblock Proc Natl Acad Sci USA 105: 4609-4614.
\input{Lipowsky2008}

\bibitem{Gennerich2006}
Gennerich A, Schild D (2006) Finite-particle tracking reveals sub-microscopic
  size changes of mitochondria during transport in mitral cell dendrites.
\newblock Phys Biol 3:45-53 3: 45-53.
\input{Gennerich2006}

\bibitem{Soppina2009}
Soppina V, Rai AK, Ramaiya AJ, Barak P, Mallik R (2009) Tug-of-war between
  dissimilar teams of microtubule motors regulates transport and fission of
  endosomes.
\newblock Proc Natl Acad Sci USA 106: 19381-19386.
\input{Soppina2009}

\bibitem{Kunwar2008}
Kunwar A, Vershinin M, Xu J, Gross SP (2008) Stepping, strain gating, and an
  unexpected force-velocity curve for multiple-motor-based transport.
\newblock Curr Biol 18: 1173-1183.
\input{Kunwar2008}

\bibitem{Kunwar2010}
Kunwar A, Mogilner A (2010) Robust transport by multiple motors with nonlinear
  force-velocity relations and stochastic load sharing.
\newblock Phys Biol 7: 016012.
\input{Kunwar2010}

\bibitem{Zhang20114}
Zhang Y (2011) Cargo transport by several motors.
\newblock Phys Rev E 83: 011909.
\input{Zhang20114}

\bibitem{Rogers2009}
Rogers AR, Driver JW, Constantinou PE, Jamison DK, Diehl MR (2009) Negative
  interference dominates collective transport of kinesin motors in the absence
  of load.
\newblock Phys Chem Chem Phys 11: 4882.
\input{Rogers2009}

\bibitem{Driver2010}
Driver J, Rogers A, Jamison D, Das R, Kolomeisky A, et~al. (2010) Coupling
  between motor proteins determines dynamic behaviors of motor protein
  assemblies.
\newblock Phys Chem Chem Phys 12: 10398-10405.
\input{Driver2010}

\bibitem{Driver2011}
Driver JW, Jamison DK, Uppulury K, Rogers AR, Kolomeisky A, et~al. (2011)
  Productive cooperation among processive motors depends inversely on their
  mechanochemical efficiency.
\newblock Biophys J 101: 386-395.
\input{Driver2011}

\bibitem{Jamison2011}
Jamison DK, Driver JW, Diehl MR (2011) Cooperative responses of multiple
  kinesins to variable and constant loads.
\newblock J Biol Chem 287: 3357-3365.
\input{Jamison2011}

\bibitem{Uppulury2012}
Uppulury K, Efremov AK, Driver JW, Jamison DK, Diehl MR, et~al. (2012) How the
  interplay between mechanical and non-mechanical interactions affect multiple
  kinesin dynamics.
\newblock J Phys Chem B 116: 8846-8855.
\input{Uppulury2012}

\bibitem{Kunwar2011}
Kunwar A, Tripathy SK, Xu J, Mattson M, Sigua R, et~al. (2011) Mechanical
  stochastic tug-of-war models cannot explain bidirectional lipid-droplet
  transport.
\newblock Proc Natl Acad Sci USA 108: 18960-18965.
\input{Kunwar2011}

\bibitem{Bouzat2012}
Bouzat S, Levi V, Bruno L (2012) Transport properties of melanosomes along
  microtubules interpreted by a tug-of-war model with loose mechanical
  coupling.
\newblock PLoS ONE 7: e43599.
\input{Bouzat2012}

\bibitem{Frank1995}
J\"{u}licher F, Prost J (1995) Cooperative molecular motors.
\newblock Phys Rev Lett 75: 2618-2621.
\input{Frank1995}

\bibitem{Badoual2002}
Badoual M, J\"{u}licher F, Prost J (2002) Bidirectional cooperative motion of
  molecular motors.
\newblock Proc Natl Acad Sci USA 99: 6696-6701.
\input{Badoual2002}

\bibitem{Adachi2007}
Adachi K, Oiwa K, Nishizaka T, Furuike S, Noji H, et~al. (2007) Coupling of
  rotation and catalysis in {$\rm F_1$-ATPase} revealed by single-molecule
  imaging and manipulation.
\newblock Cell 130: 309-321.
\input{Adachi2007}

\bibitem{Bieling2008}
Bieling P, Telley IA, Piehler J, Surrey T (2008) Processive kinesins require
  loose mechanical coupling for efficient collective motility.
\newblock EMBO Reports 19: 1121-1127.
\input{Bieling2008}

\bibitem{Mallik2009}
Mallik R, Gross SP (2009) Intracellular transport: How do motors work together?
\newblock Curr Biol 19: R416-R418.
\input{Mallik2009}

\bibitem{Brouhard2010}
Brouhard GJ (2010) Motor proteins: Kinesins influence each other through load.
\newblock Curr Biol 20: R448-R450.
\input{Brouhard2010}

\bibitem{Welte2010}
Welte MA (2010) Bidirectional transport: Matchmaking for motors.
\newblock Curr Biol 20: R410-R413.
\input{Welte2010}

\bibitem{Hendricks2010}
Hendricks AG, Perlson E, Ross JL, Schroeder HW, Tokito M, et~al. (2010) Motor
  coordination via a tug-of-war mechanism drives bidirectional vesicle
  transport.
\newblock Current Biology 20: 697-702.
\input{Hendricks2010}

\bibitem{Schroeder2010}
Schroeder HW, Mitchell C, Shuman H, Holzbaur ELF, Goldman YE (2010) Motor
  number controls cargo switching at actin-microtubule intersections in vitro.
\newblock Curr Biol 20: 687-696.
\input{Schroeder2010}

\bibitem{Leidel2012}
Leidel C, Longoria RA, Gutierrez FM, Shubeita GT (2012) Measuring molecular
  motor forces in vivo: Implications for tug-of-war models of bidirectional
  transport.
\newblock Biophys J 103: 492-500.
\input{Leidel2012}

\bibitem{Bell1978}
Bell GI (1978) Models for the specific adhesion of cells to cells.
\newblock Science 200: 618-627.
\input{Bell1978}

\bibitem{Fisher2001}
Fisher ME, Kolomeisky AB (2001) Simple mechanochemistry describes the dynamics
  of kinesin molecules.
\newblock Proc Natl Acad Sci USA 98: 7748-7753.
\input{Fisher2001}

\bibitem{Zhang20093}
Zhang Y (2009) A general two-cycle network model of molecular motors.
\newblock Physica A 383: 3465-3474.
\input{Zhang20093}

\bibitem{Zhang20112}
Zhang Y (2011) Growth and shortening of microtubules: A two-state model
  approach.
\newblock J Biol Chem 286: 39439-39449.
\input{Zhang20112}

\bibitem{Redner2001}
Redner S (2001) A Guide to First-Passage Processes.
\newblock Cambridge University Press.
\input{Redner2001}

\bibitem{Zhang20113}
Zhang Y (2011) Periodic one-dimensional hopping model with transitions between
  nonadjacent states.
\newblock Phys Rev E 84: 031104.
\input{Zhang20113}

\end{thebibliography}
\end{document}